\documentclass[10pt,a4paper]{article}
\usepackage[utf8]{inputenc}
\usepackage{amsmath}
\usepackage{amsfonts}
\usepackage{amssymb}
\author{D. Rayneau-Kirkhope$^{1,2}$, Chengzhao Zhang$^{1,3}$, Louis Theran$^{1,4}$ and\\ Marcelo A. Dias$^{1,5}$}

\usepackage{graphicx,color,verbatim}
\title{Analytic analysis of auxetic metamaterials through analogy with rigid link systems}
\begin{document}
\maketitle
\begin{center}
$^{1}$Aalto Science Institute, Aalto University, FI-02150 Espoo, Finland\\
$^{2}$Department of Applied Physics, Aalto University, FI-02150 Espoo, Finland\\
$^{3}$Department of Mathematics, Massachusetts Institute of Technology, USA\\
$^{4}$School of Mathematics and Statistics, University of St Andrews, St Andrews KY16 9SS, Scotland, UK\\
$^{5}$Department of Physics and Astronomy, James Madison University, Harrisonburg, VA 22807, USA
\end{center}
\begin{abstract}
Recent progress in advanced additive manufacturing techniques has stimulated the growth of the field of mechanical metamaterials. 
One area particular interest in this subject is the creation of auxetic material properties through elastic instability.
This paper focuses on a novel methodology in the analysis of auxetic metamaterials through analogy with rigid link lattice systems. 
Our analytic methodology gives extremely good agreement with finite element simulations for both the onset of elastic instability and post-buckling behaviour including Poisson's ratio. 
The insight into the relationships between mechanisms within lattices and their mechanical behaviour has the potential to guide the rational design of lattice based metamaterials.
\end{abstract}

\section{Introduction}

With the close control of a materials internal geometry, comes the potential to design a particular response to load into the architecture of a material. Such structures, where geometry rather than material properties govern the macroscopic response of a solid, are termed ``mechanical metamaterials'' \cite{Meta_materials1, Meta_materials2}. Perhaps the most well-known class of these kind of materials are those with an auxetic reponse to external load \cite{Auxetic_1,Auxetic_2}. Such structures, when compressed (stretched), contract (expand) in the direction perpendicular to the applied load.  

Auxetic behaviour has been observed in layered ceramics \cite{Layered_ceramics}, re-entrant foams \cite{Auxetic_1}, origami \cite{Origami} and kirigami \cite{Kirigami_1, Kirigami_2} geometries, structures with rotating elements \cite{Rotating}, dimpled sheets \cite{Dimpled} and other carefully designed architectures. 
Some of these geometries are examples of an emerging paradigm whereby the use of elastic instability is viewed as a route to new functionality rather than a mode of failure \cite{Elastic_instability1, Elastic_instability2}. 
This concept has been notably used in the creation of materials endowed with negative poisson ratio, where spontaneous symmetry breaking associated with buckling creates an ordered collapse and an auxetic response beyond some threshold value of loading \cite{Buckling_aux1, Buckling_aux2, Buckling_aux3}. 
Alongside auxetic response, metamaterials exhibiting frustrated mechanics \cite{Frustrated} and hysteretic \cite{Hysteretic} behaviours have been designed and fabricated using similar buckling induced motifs.
More broadly, the utilisation of elastic instabilities in soft materials resulting in novel mechanical behaviour has been bought under the umbrella term ``designer matter'' \cite{Designer}. 
This emerging field is notable for embracing truly interdisciplinary work, encompassing fields including engineering, physics, architecture and increasingly applied mathematics.

In this paper, we establish a fundamental connection between the buckling-induced auxetic behaviour of a lattice made up of a soft material and the mechanisms (``floppy modes'') designed into a lattice of rigid links. 
We import the knowledge of the mechanisms of the rigid link lattices and apply it to analytically establish the buckling and postbuckling behaviour of complex continuum lattice systems. 
In order to facilitate this analogy, we utilise a specific void shape allowing the straightforward calculation of the appropriate values for the torsional springs in the system. 
We note however, that the underlying mechanisms within the continuum lattice are relatively insensitive to the nature of the voids \cite{Hole_shape}, thus we hypothesise that the rigid-link analysis proposed here elucidates the fundamental mechanisms present within a wide range of elastic instabilities in lattice based materials.
In this paper, we choose to focus on the auxetic properties of lattice based materials, however, the methodology presented here can be applied to more complex lattice problems, including mechanical hysteresis behaviour and/or frustrated mechanics \cite{Frustrated, Hysteretic}. 
Such behaviour can be studied through the introduction of more complex mechanisms which permit further degrees of freedom and/or constraints within the rigid-link lattice. 
It is also noted that while good agreement between applied load and resultant deformation is observed here, the model is unsuitable to model displacement controlled relationships (with the exception of the ``far-from-thershold'' regime) due the the rigid link nature of the links considered here.

This paper focuses on analytic solutions for the rigid link lattice system for the onset of elastic instability and its associated eigenmode, the postbuckling stiffness and associated properties such as the Poisson's ratio of the system (found to be negative for certain regions of parameter space). 
We present close agreement for all of these properties with an analogous continuum system investigated through finite element methods. 
We propose that the use of rigid link methods can be a powerful tool in the analysis of elastic systems providing insight into the fundamental mechanisms/modes present within the lattice and reducing the computational power required for such investigations.

\begin{figure}
\begin{center}
\resizebox{7cm}{!}{\begingroup%
  \makeatletter%
  \providecommand\color[2][]{%
    \errmessage{(Inkscape) Color is used for the text in Inkscape, but the package 'color.sty' is not loaded}%
    \renewcommand\color[2][]{}%
  }%
  \providecommand\transparent[1]{%
    \errmessage{(Inkscape) Transparency is used (non-zero) for the text in Inkscape, but the package 'transparent.sty' is not loaded}%
    \renewcommand\transparent[1]{}%
  }%
  \providecommand\rotatebox[2]{#2}%
  \ifx\svgwidth\undefined%
    \setlength{\unitlength}{217.83840332bp}%
    \ifx\svgscale\undefined%
      \relax%
    \else%
      \setlength{\unitlength}{\unitlength * \real{\svgscale}}%
    \fi%
  \else%
    \setlength{\unitlength}{\svgwidth}%
  \fi%
  \global\let\svgwidth\undefined%
  \global\let\svgscale\undefined%
  \makeatother%
  \begin{picture}(1,0.69411771)%
    \put(0,0){\includegraphics[width=\unitlength,page=1]{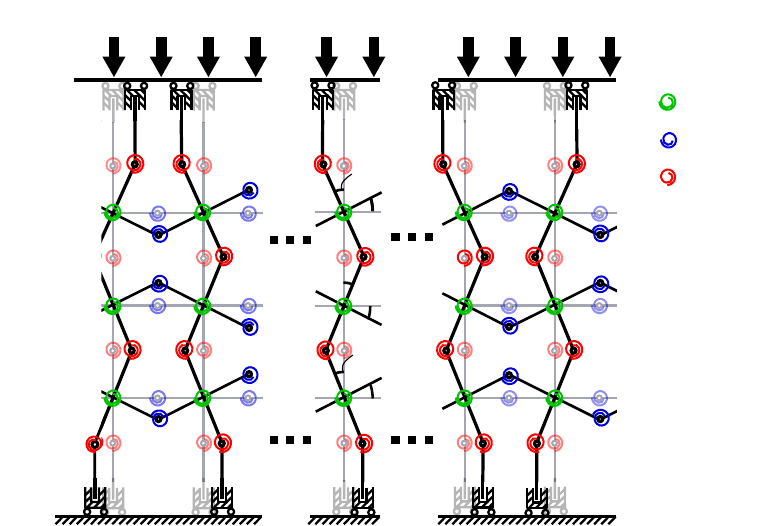}}%
    \put(0.50674697,0.26508434){\color[rgb]{0,0,0}\makebox(0,0)[lb]{\smash{$\eta_{i,2}$}}}%
    \put(0.50674697,0.16842356){\color[rgb]{0,0,0}\makebox(0,0)[lb]{\smash{$\eta_{i,1}$}}}%
    \put(0.50674697,0.41410302){\color[rgb]{0,0,0}\makebox(0,0)[lb]{\smash{$\eta_{i,3}$}}}%
    \put(0.46647167,0.45437835){\color[rgb]{0,0,0}\makebox(0,0)[lb]{\smash{$\theta_{i,3}$}}}%
    \put(0.37857614,0.33213192){\color[rgb]{0,0,0}\makebox(0,0)[lb]{\smash{$\theta_{i,2}$}}}%
    \put(0.46647167,0.20869885){\color[rgb]{0,0,0}\makebox(0,0)[lb]{\smash{$\theta_{i,1}$}}}%
    \put(0.90579251,0.55080978){\color[rgb]{0,0,0}\makebox(0,0)[lb]{\smash{$\tau$}}}%
    \put(0.90593611,0.49925068){\color[rgb]{0,0,0}\makebox(0,0)[lb]{\smash{$\kappa_{1}$}}}%
    \put(0.90603634,0.449701){\color[rgb]{0,0,0}\makebox(0,0)[lb]{\smash{$\kappa_{2}$}}}%
    \put(0.44897623,0.64882812){\color[rgb]{0,0,0}\makebox(0,0)[lb]{\smash{$F$}}}%
  \end{picture}%
\endgroup%
}
\resizebox{6cm}{!}{\begingroup%
  \makeatletter%
  \providecommand\color[2][]{%
    \errmessage{(Inkscape) Color is used for the text in Inkscape, but the package 'color.sty' is not loaded}%
    \renewcommand\color[2][]{}%
  }%
  \providecommand\transparent[1]{%
    \errmessage{(Inkscape) Transparency is used (non-zero) for the text in Inkscape, but the package 'transparent.sty' is not loaded}%
    \renewcommand\transparent[1]{}%
  }%
  \providecommand\rotatebox[2]{#2}%
  \ifx\svgwidth\undefined%
    \setlength{\unitlength}{239.99999279bp}%
    \ifx\svgscale\undefined%
      \relax%
    \else%
      \setlength{\unitlength}{\unitlength * \real{\svgscale}}%
    \fi%
  \else%
    \setlength{\unitlength}{\svgwidth}%
  \fi%
  \global\let\svgwidth\undefined%
  \global\let\svgscale\undefined%
  \makeatother%
  \begin{picture}(1,0.83550577)%
    \put(0,0){\includegraphics[width=\unitlength,page=1]{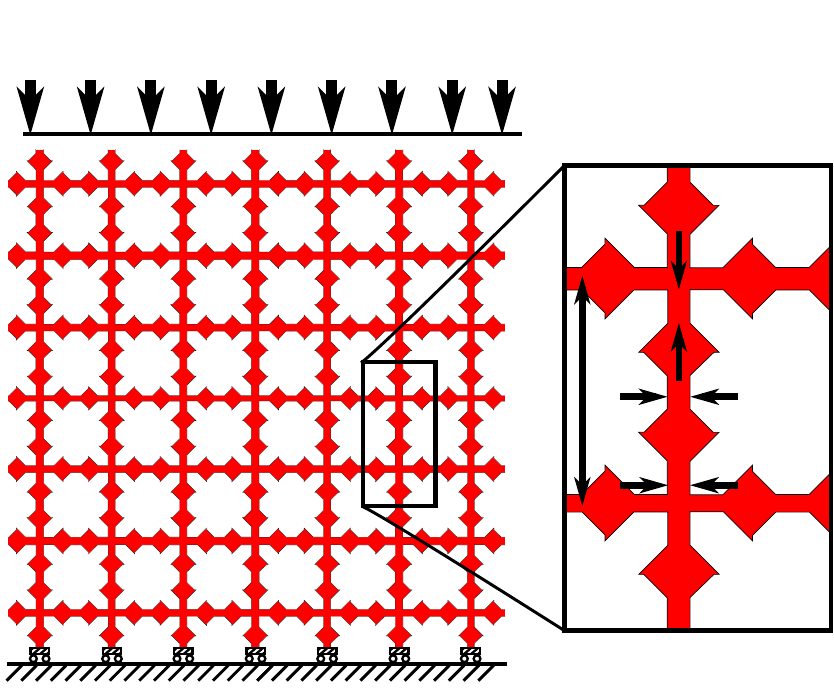}}%
    \put(1.72373298,0.63100439){\color[rgb]{0,0,0}\makebox(0,0)[lb]{\smash{}}}%
    \put(0,0){\includegraphics[width=\unitlength,page=2]{lattice3.pdf}}%
    \put(0.83800257,0.533417){\color[rgb]{0,0,0}\makebox(0,0)[lb]{\smash{$l_c$}}}%
    \put(0.85018824,0.37893925){\color[rgb]{0,0,0}\makebox(0,0)[lb]{\smash{$(1-a)L$}}}%
    \put(0.85446741,0.27671371){\color[rgb]{0,0,0}\makebox(0,0)[lb]{\smash{$(1-b)L$}}}%
    \put(0.70933924,0.39630538){\color[rgb]{0,0,0}\makebox(0,0)[lb]{\smash{$L$}}}%
    \put(0.27345231,0.7531631){\color[rgb]{0,0,0}\makebox(0,0)[lb]{\smash{$F$}}}%
  \end{picture}%
\endgroup%
}
\caption{\footnotesize Left: The rigid link lattice is comprised of infinitely stiff members connected by rotational springs. 
At the hinges, between two neighbouring links, springs of stiffnesses $\kappa_1$ and $\kappa_2$ are placed. Where two links overlap, they are assumed to be pin jointed, rotation relative to one another is permitted at the expense of deforming a torsional spring of stiffness $\tau$. Right: An analogous continuum lattice is shown, where red indicates a soft elastomer while white shows the voids. The boundary conditions are shown schematically in the diagram -- on the horizontal boundaries the free ends of the lattice are assumed to translate in the $y$ direction together, while translations in the $x$ direction are free. On the vertical boundaries translation of the free ends in both the $x$ and $y$ direction are permitted. On both boundaries, rotations are not permitted. \label{Systems}}
\end{center}
\end{figure}

\section{Lattice stability}

The deformation of the rigid link lattice presented in Fig.~(\ref{Systems}, left) is parameterised by a set of angles $\{\theta_{i,j}\}$ denoting the rotation of the initially vertical elements, while another set of angles $\{\eta_{i,j}\}$ is used to represent the rotations of the initially horizontal elements. The indices $i$ and $j$ take integer values and denote the position of the element: element $(i,j)$ has its centre initially positioned at $(iL,jL)$ where $L$ is the length of the links. There are no horizontal element on the upper and lower boundaries ($j=0,N_y$). 
The energy of any deformation described by the parameters $\{\theta_{i,j}, \eta_{i,j}\}$ can be calculated as,
\begin{eqnarray}
U &=& \sum_{i=1}^{N_x}\sum_{j=0}^{N_y} \frac{\kappa_1\left(1+\delta_{j,1} + \delta_{j,N_y}\right)}{2}\left(\theta_{i,j+1} - \theta_{i,j}\right)^2 + \sum_{i=1}^{N_x}\sum_{j=1}^{N_y} \frac{\tau}{2}\left(\theta_{i,j} - \eta_{i,j}\right)^2\nonumber\\ &&+ \sum_{i=1}^{N_x-1}\sum_{j=1}^{N_y} \frac{\kappa_2}{2}\left(\eta_{i+1,j} - \eta_{i,j}\right)^2 -  \sum_{i=1}^{N_x}F_i l \sum_{j=1}^{N_y} \left(1 - \cos\theta_{i,j}\right),\label{energy}
\end{eqnarray}
where $\kappa_1$, $\kappa_2$ and $\tau$ are the stiffnesses of the torsional springs in the system (springs with strength $\kappa_{1}$ ($\kappa_2$) penalise rotation of initially vertical (horizontal) elements relative to their neighbours while $\tau$ springs penalise relative rotation of the vertical and horizontal elements whose centres remain coincident).
The term proportional to $\kappa_1$ contains the Kronecker delta function $\delta_{i,j}$ to account for our choice of boundary constraints. The last term in Eq.~\eqref{energy} is the external work done on the system by the force $F_i$ (at each $i$-th vertical element).
Due to the connectivity of the lattice and fixed length of elements, we impose a set of constraints. 
Working to first order in $\{\theta_{i,j},\eta_{i,j}\}$, it can be shown that for a deformation to be compatible with the connectivity of the lattice, the distance between the locations of $\eta_{i,j}$ and $\eta_{i+1,j}$ takes a constant value for all $j$. Furthermore it is required that
\begin{equation}
\eta_{i,j} = -\eta_{i+1,j}.
\end{equation}
We set the boundary conditions of the system to be, 
\begin{eqnarray}
\theta_{i,0} = -\theta_{i,1},\\
\theta_{i,N_y+1} = - \theta_{i,Ny}.
\end{eqnarray}
\subsection{Symmetry relations}
Following some of the authors previous work~\cite{Symmetry}, it can be assumed that the energy of the system is minimised when one of two symmetry relations are assumed:
\begin{align}
\theta_{i,j} = \theta_{i+1,j},\\
\theta_{i,j} = -\theta_{i+1,j}.
\end{align}
These modes will be referred to as translationally symmetric and mirror symmetric modes, respectively. In the following subsections, we establish the buckling behaviour of the system subject to these two possible symmetry relations. 

\subsubsection{Translational symmetry}
Utilising the symmetry relationships presented above, the energy of a given deformation can be greatly simplified. In the case of translational symmetry ($\theta_{i,j} = \theta_{i+1,j}$) we see that energy minimisation with respect to $\eta_{i,j}$ enforces that 
\begin{equation}
\eta_{i,j} = 0 \quad \forall \quad i,j.
\end{equation}
Working to first order in $\theta_{i,j}$, direct calculation shows that the minimum energy configuration exists when 
\begin{equation}
(2\kappa_1 + \tau -Fl)\theta_{n,m} - \kappa_1(\theta_{n,m+1} + \theta_{n,m-1}) +(\delta_{m,1}+\delta_{m,Ny})2\kappa_1 = 0 \label{sym_Emin}
\end{equation}
is satisfied for all values of $n$ and $m$. This requirement can be rewritten in matrix form, 
\begin{equation}
\mathbf{A}\Theta  = \mathbf{0}
\end{equation}
where $\Theta = (\theta_{i,1}, \theta_{i,2}, ... ,\theta_{n,N_y})^T$. It is noted that in matrix form the first two terms of Eq.~\ref{sym_Emin} create a tridiagonal symmetric Toeplitz matrix, the remaining term mean that the full expression for $\mathbf{A}$ deviates slightly from this form. Buckling of the system into a mode with translational symmetry will occur if the loading on the system is sufficient to create (atleast) one zero eigenvalue of $\mathbf{A}$. For suitably large values of $N_y$, neglecting the last term in Eq.~(\ref{sym_Emin}) yields a good approximation to the system (for physically relevant parameters here, $N_y > 5$ is sufficient). This approximation allows for analytic calculation of the eigenvalues/vectors. We find that buckling of the system into a mode with translational symmetry will occur if the loading on the system exceeds the threshold
\begin{equation}
F_\text{min} = \frac{2\kappa_1 + \tau - 2\kappa_1\cos\left(\frac{\pi}{N_y+1}\right)}{l},\label{F_l}
\end{equation} 
and the associated mode is found to be 
\begin{equation}
\Theta_i = A\sin\left(\frac{i\pi}{n+1}\right).\label{mode_sym}
\end{equation}
For small values of $N_y$, the eigenmode of this system can be obtained numerically. 

\subsubsection{Mirror symmetry}
The second symmetry between neighbouring columns we consider here is that of mirror symmetry ($\theta_{i,j} = -\theta_{i+1,j}$). Due to the restriction that the mid-point of vertical and horizontal bars are coincident throughout the deformation process, it can be shown that,
\begin{equation}
\theta_{i,j} =\pm \theta \label{mode_asym_a}
\end{equation}
for some value of $\theta$ and that the sign of any rotation in the lattice is the opposite of its nearest neighbours.  
Furthermore through the minimisation of energy with respect to $\eta$, we can also derive the an expression for the rotation of initially vertical elements:
\begin{equation}
\eta_{i,j} =  \gamma(N_x,\tau,\kappa_2) \theta_{i,j}. \label{mode_asym_b}
\end{equation}
The energy of the whole system for a given deformation can then be expressed as
\begin{equation}
 U =  \Omega \theta^2 -  F_ilN_xN_y(1-\cos\theta)
\end{equation}
for some $\Omega$ which depends on parameters describing the lattice. 
Working to first order in $\theta$, we can thus establish that the minimum energy configuration corresponds to non-zero values of $\theta$ (buckled configurations), provided $F$ is above 
\begin{equation}
F_\text{min} = \frac{2\Omega }{N_xN_yl}.\label{F_s}
\end{equation}

\section{Rigid link as a continuum approximation}

If we now consider a continuum lattice structure, as shown in figure \ref{Systems} (right), the slender beams within the framework serve as hinge points when the lattice deforms beyond the buckling threshold. The resistance to bending of these slender beams is easily obtainable and, thus, we are able to calculate the effective values of $\kappa_1$, $\kappa_2$ and $\tau$ of the continuum lattice. From standard beam theory~\cite{Timoshenko}, it can be found that the appropriate stiffnesses of these springs are given by 
\begin{align}
\kappa_1 = \kappa_2 = \frac{EI_\kappa}{l_c}\label{kappa}\\
\tau = \frac{2EI_\tau}{l_c} \label{tau},
\end{align}
where $E$ is the Young's Modulus of the material, $I_\kappa$ and $I_\tau$ are the second moment of area of the slender elements making up the $\kappa_1$, $\kappa_2$ and $\tau$ springs and $l_c$ is the length of the slender elements. 
Therefore, using the expressions for buckling of the rigid link lattice found in the previous section, we can predict the buckling load of the more complex continuum lattice system.
 
\subsection{Linear stability}
\begin{figure}
\begin{center}
\includegraphics[scale=0.2]{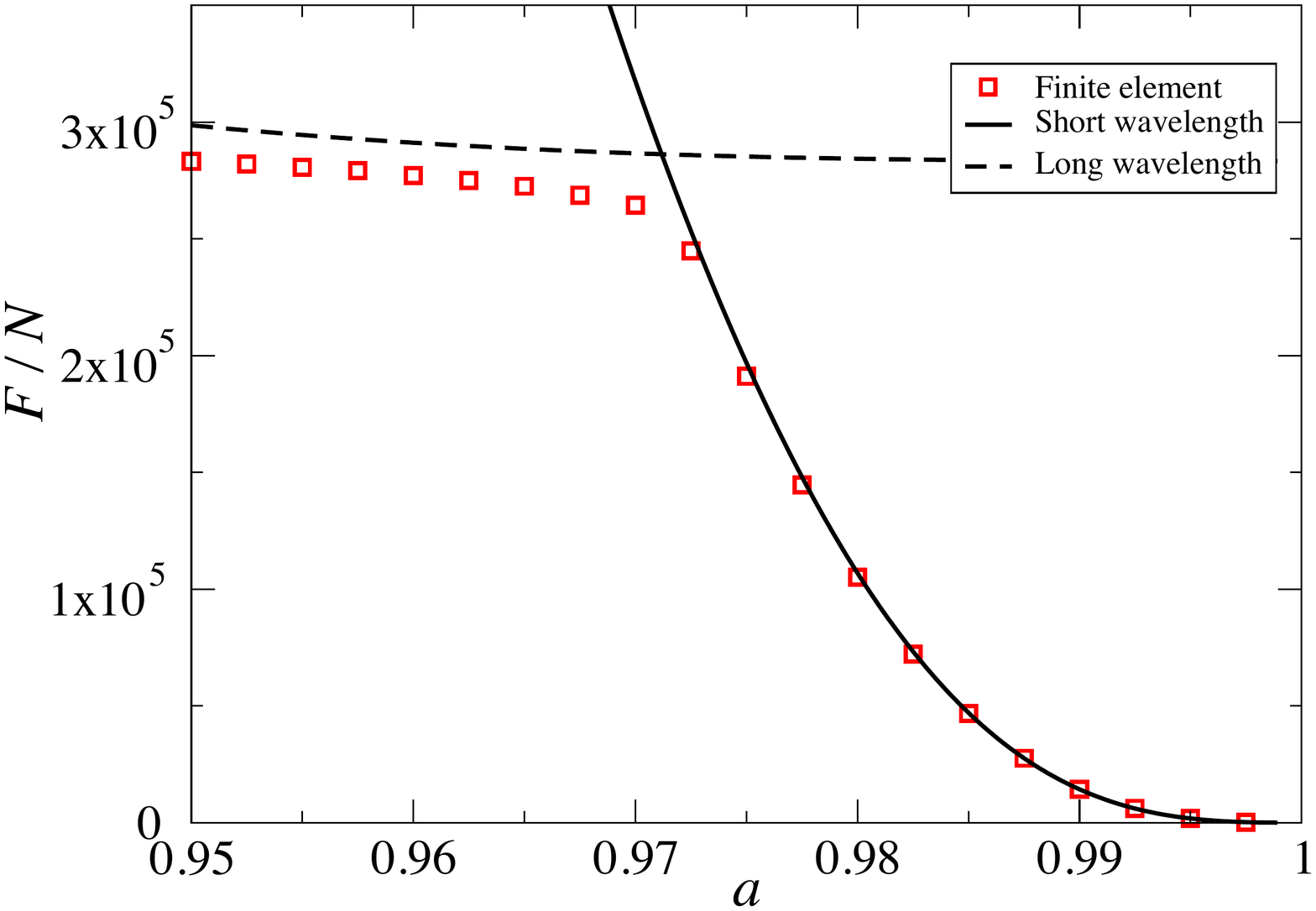}
\includegraphics[scale=0.2]{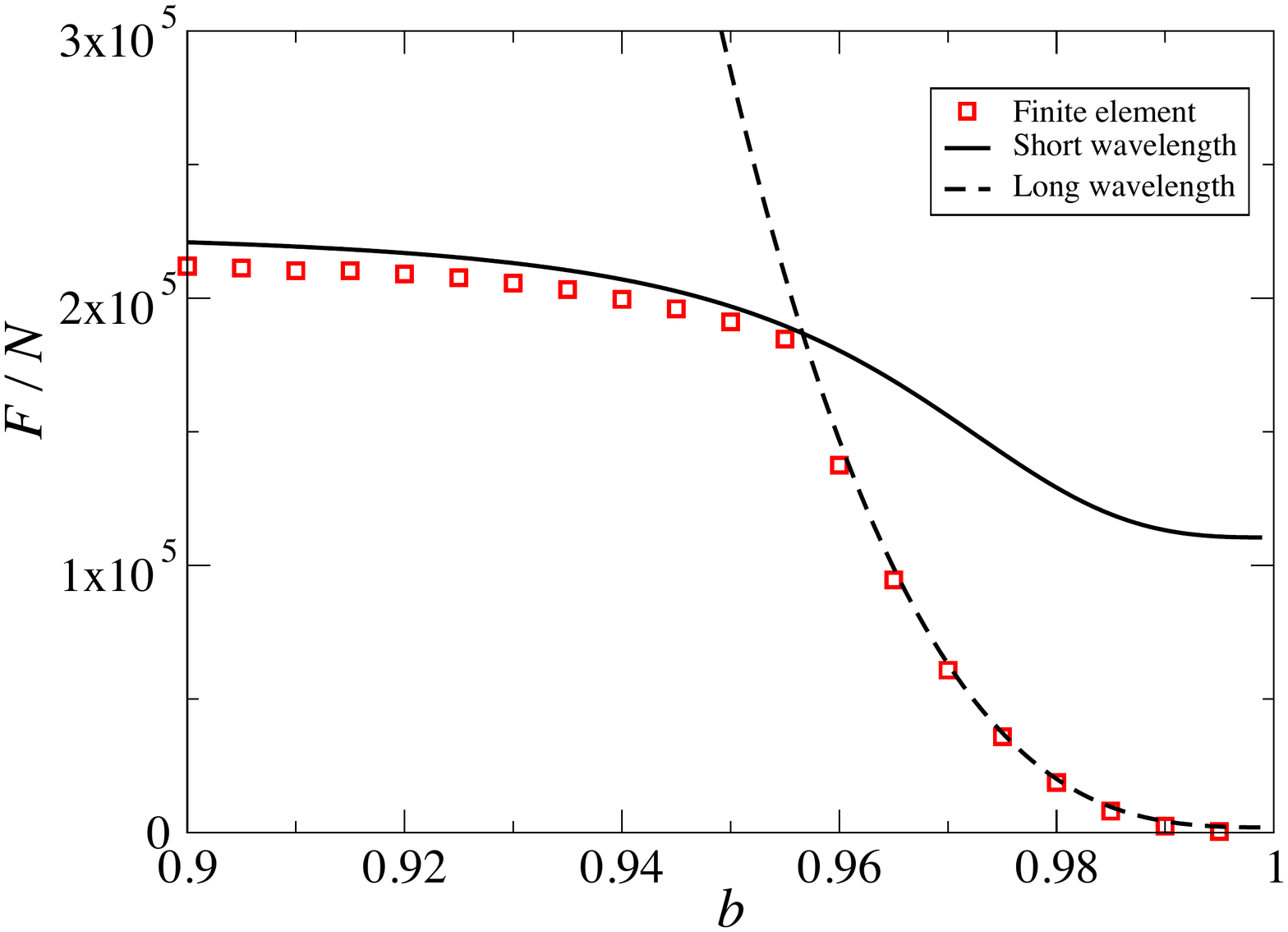}
\end{center}
\caption{Comparison between finite element simulations and the analytic results derived in Eqs.~\ref{F_l} \& \ref{F_s}. Both plots show results found for parameters $N_x = 10, N_y = 9$, $l_c = 3.75$mm and $L = 20$mm, $\tau$ and $\kappa_n$ are given in Eq.~(\ref{kappa} \& \ref{tau}). Left: failure load for constant $b = 0.95$. Right: Failure load for constant $a = 0.975$.\label{linear}}
\end{figure}

\begin{figure}
\begin{center}
\includegraphics[scale=0.2]{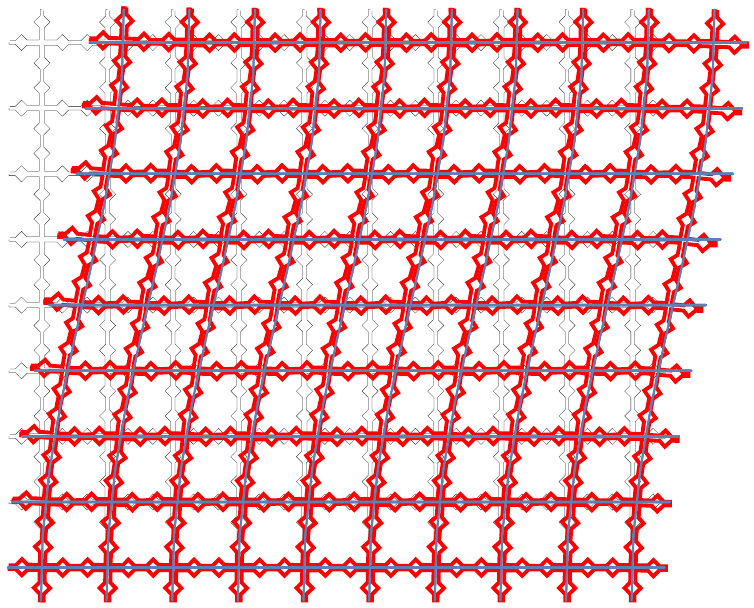}
\includegraphics[scale=0.185]{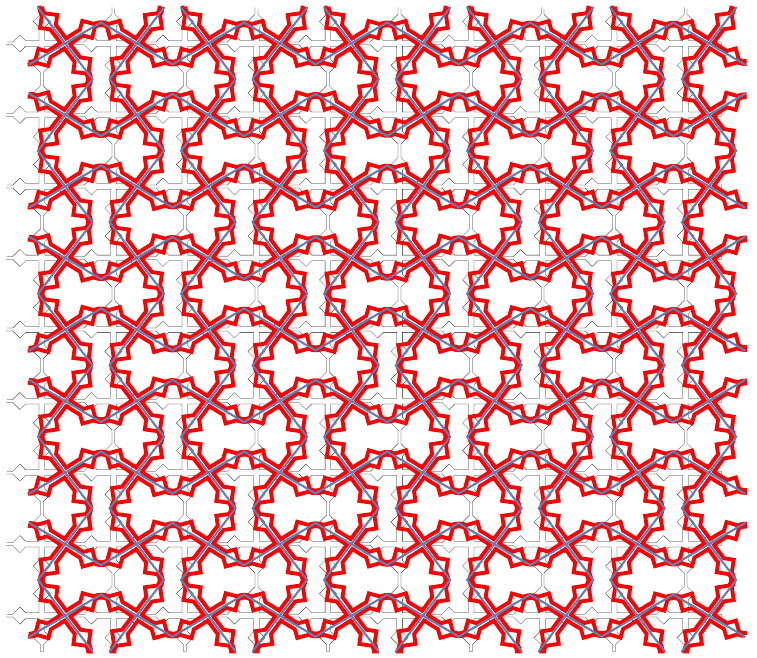}
\end{center}
\caption{Good agreement is found for the linear buckling mode as predicted by the rigid link method (grey line) and the finite element work (red outline). Left showing the mode with translational symmetry, right with mirror symmetry. Here, the parameters used are $N_x = 10, N_y = 9, L= 20$mm, $l_c = 3.75$mm, $b = 0.92$ and $a = 0.945$ (left) and $a = 0.96$ (right).\label{mode_diagrams}}
\end{figure}

We note that the minimum value of the two buckling loads of the lattice (corresponding translational or mirror symmetry) will be the physically relevant mode. 
Thus, we are able to make predictions about the regions of the parameter space describing the continuum lattice and how they will yield modes that correspond to  either of the two symmetries mentioned in the previous section. 
For a set values of $N_x = 10, N_y = 9$, $l_c = 3.75$mm and $L = 20$mm, the thicknesses of the slender beam elements are varied through the dimensionless parameters $a$ and $b$ (see figure \ref{Systems} (right)). This in turn sets values of $\kappa$ and $\tau$ in the rigid link lattice through Eqs.~(\ref{kappa} \& \ref{tau}). 
\begin{figure}
\begin{center}
\includegraphics[scale=0.25]{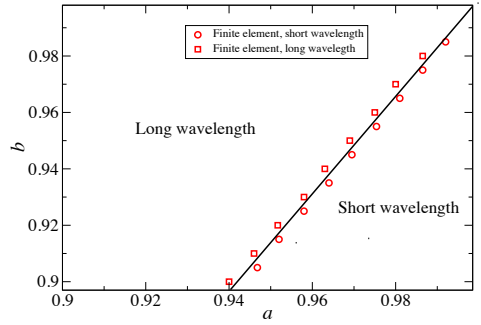}
\end{center}
\caption{The predicted failure mode for the continuous system for varied values of $a$ and $b$. The black line separates the two modes as predicted by the rigid link model and red squares and circles are finite element confirmation of the boundary showing long and short wavelength modes respectively. \label{phases}}
\end{figure} 
Predictions of which symmetry mode is present within the deformed structure are then found through the rigid link methodology. The results are shown in a phase diagram, for the space $(a,b)$, where a boundary between the translationally symmetric mode (long wavelength) and mirror symmetric mode (short wavelength) is analytically drawn (see figure \ref{phases}). 
Good agreement between these predictions and the modes computed in finite element work is observed.
Furthermore, we can predict the buckling load of the continuum structure through substituting Eqs.~(\ref{kappa} \& \ref{tau}) into Eqs.~(\ref{F_l} \& \ref{F_s}). 
The agreement in buckling load between the rigid link method and the finite element simulations is shown in figure \ref{linear}, where increasingly better quantitative match is found for more slender beam elements.
It is also noted that the linear buckling modes, as predicted in Eqs.~(\ref{mode_sym}, \ref{mode_asym_a} \& \ref{mode_asym_b}) agrees well with the modes found in the linear analysis through finite element studies, as shown in figure \ref{mode_diagrams}. 

\newpage

\subsection{Postbuckling}
\begin{figure}
\begin{center}
\resizebox{6cm}{!}{\begingroup%
  \makeatletter%
  \providecommand\color[2][]{%
    \errmessage{(Inkscape) Color is used for the text in Inkscape, but the package 'color.sty' is not loaded}%
    \renewcommand\color[2][]{}%
  }%
  \providecommand\transparent[1]{%
    \errmessage{(Inkscape) Transparency is used (non-zero) for the text in Inkscape, but the package 'transparent.sty' is not loaded}%
    \renewcommand\transparent[1]{}%
  }%
  \providecommand\rotatebox[2]{#2}%
  \ifx\svgwidth\undefined%
    \setlength{\unitlength}{264bp}%
    \ifx\svgscale\undefined%
      \relax%
    \else%
      \setlength{\unitlength}{\unitlength * \real{\svgscale}}%
    \fi%
  \else%
    \setlength{\unitlength}{\svgwidth}%
  \fi%
  \global\let\svgwidth\undefined%
  \global\let\svgscale\undefined%
  \makeatother%
  \begin{picture}(1,0.70290019)%
    \put(0,0){\includegraphics[width=\unitlength,page=1]{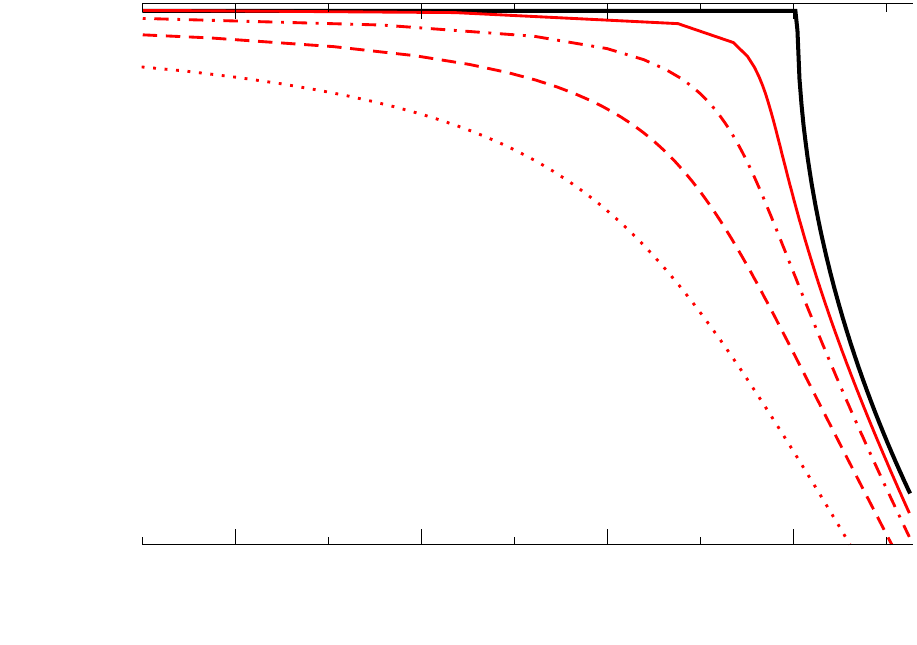}}%
    \put(0.20910172,0.06433805){\color[rgb]{0,0,0}\makebox(0,0)[lb]{\smash{-0.03}}}%
    \put(0.41131023,0.06433805){\color[rgb]{0,0,0}\makebox(0,0)[lb]{\smash{-0.02}}}%
    \put(0.61579357,0.06433805){\color[rgb]{0,0,0}\makebox(0,0)[lb]{\smash{-0.01}}}%
    \put(0.85758996,0.06433805){\color[rgb]{0,0,0}\makebox(0,0)[lb]{\smash{0}}}%
    \put(0.45688009,0.00735786){\color[rgb]{0,0,0}\makebox(0,0)[lb]{\smash{$\frac{F-F_{\mbox{min}}}{F_{\mbox{min}}}$}}}%
    \put(0,0){\includegraphics[width=\unitlength,page=2]{Sym_bif.pdf}}%
    \put(0.04465574,0.13585801){\color[rgb]{0,0,0}\makebox(0,0)[lb]{\smash{-0.07}}}%
    \put(0.0439817,0.21345555){\color[rgb]{0,0,0}\makebox(0,0)[lb]{\smash{-0.06}}}%
    \put(0.04499275,0.29088455){\color[rgb]{0,0,0}\makebox(0,0)[lb]{\smash{-0.05}}}%
    \put(0.04364469,0.36873482){\color[rgb]{0,0,0}\makebox(0,0)[lb]{\smash{-0.04}}}%
    \put(0.04532977,0.44633236){\color[rgb]{0,0,0}\makebox(0,0)[lb]{\smash{-0.03}}}%
    \put(0.04364469,0.52392987){\color[rgb]{0,0,0}\makebox(0,0)[lb]{\smash{-0.02}}}%
    \put(0.04667782,0.60152739){\color[rgb]{0,0,0}\makebox(0,0)[lb]{\smash{-0.01}}}%
    \put(0.09984181,0.6791249){\color[rgb]{0,0,0}\makebox(0,0)[lb]{\smash{0}}}%
    \put(0.03199563,0.33975161){\color[rgb]{0,0,0}\rotatebox{90}{\makebox(0,0)[lb]{\smash{$\theta_1$}}}}%
    \put(0,0){\includegraphics[width=\unitlength,page=3]{Sym_bif.pdf}}%
    \put(0.30084252,0.33137663){\color[rgb]{0,0,0}\makebox(0,0)[lb]{\smash{Rigid link}}}%
    \put(0,0){\includegraphics[width=\unitlength,page=4]{Sym_bif.pdf}}%
    \put(0.30084252,0.2893454){\color[rgb]{0,0,0}\makebox(0,0)[lb]{\smash{FEM, $\zeta = 2.4\times 10^{-4}$}}}%
    \put(0,0){\includegraphics[width=\unitlength,page=5]{Sym_bif.pdf}}%
    \put(0.30084252,0.24832522){\color[rgb]{0,0,0}\makebox(0,0)[lb]{\smash{FEM, $\zeta = 1.2\times 10^{-4}$}}}%
    \put(0,0){\includegraphics[width=\unitlength,page=6]{Sym_bif.pdf}}%
    \put(0.30084252,0.20730504){\color[rgb]{0,0,0}\makebox(0,0)[lb]{\smash{FEM, $\zeta = 6\times 10^{-5}$}}}%
    \put(0,0){\includegraphics[width=\unitlength,page=7]{Sym_bif.pdf}}%
    \put(0.30084252,0.16628484){\color[rgb]{0,0,0}\makebox(0,0)[lb]{\smash{FEM, $\zeta = 3\times 10^{-5}$}}}%
    \put(0,0){\includegraphics[width=\unitlength,page=8]{Sym_bif.pdf}}%
  \end{picture}%
\endgroup%
}
\resizebox{6cm}{!}{\begingroup%
  \makeatletter%
  \providecommand\color[2][]{%
    \errmessage{(Inkscape) Color is used for the text in Inkscape, but the package 'color.sty' is not loaded}%
    \renewcommand\color[2][]{}%
  }%
  \providecommand\transparent[1]{%
    \errmessage{(Inkscape) Transparency is used (non-zero) for the text in Inkscape, but the package 'transparent.sty' is not loaded}%
    \renewcommand\transparent[1]{}%
  }%
  \providecommand\rotatebox[2]{#2}%
  \ifx\svgwidth\undefined%
    \setlength{\unitlength}{280.0000024bp}%
    \ifx\svgscale\undefined%
      \relax%
    \else%
      \setlength{\unitlength}{\unitlength * \real{\svgscale}}%
    \fi%
  \else%
    \setlength{\unitlength}{\svgwidth}%
  \fi%
  \global\let\svgwidth\undefined%
  \global\let\svgscale\undefined%
  \makeatother%
  \begin{picture}(1,0.68447558)%
    \put(0,0){\includegraphics[width=\unitlength,page=1]{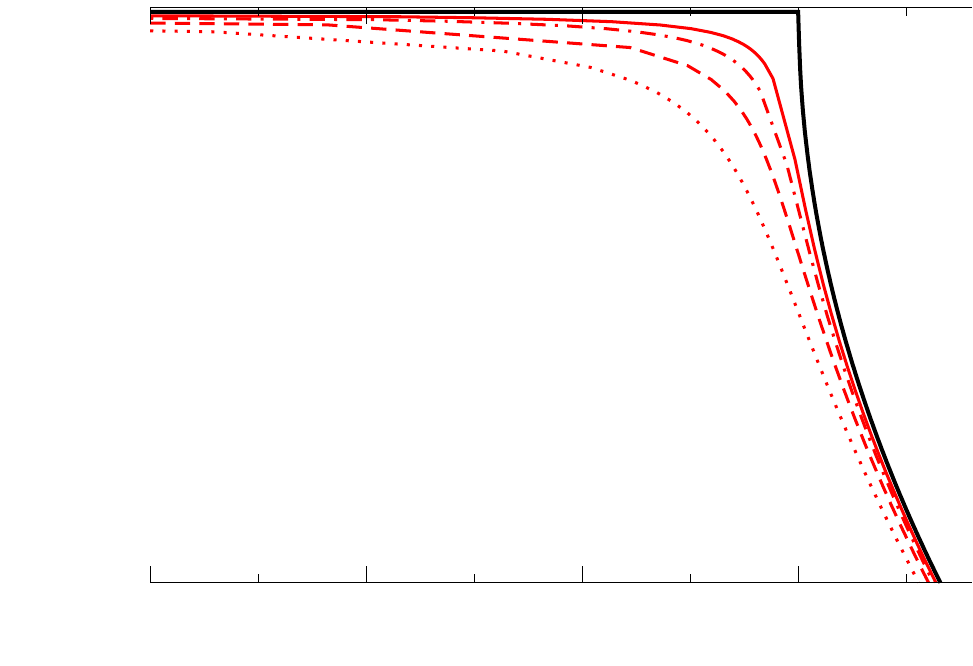}}%
    \put(0.10990001,0.05248159){\color[rgb]{0,0,0}\makebox(0,0)[lb]{\smash{-0.015}}}%
    \put(0.34179147,0.05281948){\color[rgb]{0,0,0}\makebox(0,0)[lb]{\smash{-0.01}}}%
    \put(0.55450649,0.05248159){\color[rgb]{0,0,0}\makebox(0,0)[lb]{\smash{-0.005}}}%
    \put(0.81309295,0.05281948){\color[rgb]{0,0,0}\makebox(0,0)[lb]{\smash{0}}}%
    \put(0,0){\includegraphics[width=\unitlength,page=2]{ASym_bif.pdf}}%
    \put(0.0500416,0.15740296){\color[rgb]{0,0,0}\makebox(0,0)[lb]{\smash{-0.12}}}%
    \put(0.07090761,0.24128938){\color[rgb]{0,0,0}\makebox(0,0)[lb]{\smash{-0.1}}}%
    \put(0.05105534,0.3251758){\color[rgb]{0,0,0}\makebox(0,0)[lb]{\smash{-0.08}}}%
    \put(0.05037952,0.40897775){\color[rgb]{0,0,0}\makebox(0,0)[lb]{\smash{-0.06}}}%
    \put(0.0500416,0.49286417){\color[rgb]{0,0,0}\makebox(0,0)[lb]{\smash{-0.04}}}%
    \put(0.0500416,0.57675058){\color[rgb]{0,0,0}\makebox(0,0)[lb]{\smash{-0.02}}}%
    \put(0.10638829,0.660637){\color[rgb]{0,0,0}\makebox(0,0)[lb]{\smash{0}}}%
    \put(0,0){\includegraphics[width=\unitlength,page=3]{ASym_bif.pdf}}%
    \put(0.35298545,0.29481197){\color[rgb]{0,0,0}\makebox(0,0)[lb]{\smash{Rigid link}}}%
    \put(0,0){\includegraphics[width=\unitlength,page=4]{ASym_bif.pdf}}%
    \put(0.35298545,0.25339376){\color[rgb]{0,0,0}\makebox(0,0)[lb]{\smash{FEM, $\zeta=6.25\times 10^{-5}$}}}%
    \put(0,0){\includegraphics[width=\unitlength,page=5]{ASym_bif.pdf}}%
    \put(0.35298545,0.21695978){\color[rgb]{0,0,0}\makebox(0,0)[lb]{\smash{FEM, $\zeta=3.13\times 10^{-5}$}}}%
    \put(0,0){\includegraphics[width=\unitlength,page=6]{ASym_bif.pdf}}%
    \put(0.35298545,0.1805258){\color[rgb]{0,0,0}\makebox(0,0)[lb]{\smash{FEM, $\zeta=1.56\times10^{-5}$}}}%
    \put(0,0){\includegraphics[width=\unitlength,page=7]{ASym_bif.pdf}}%
    \put(0.35298545,0.1440918){\color[rgb]{0,0,0}\makebox(0,0)[lb]{\smash{FEM, $\zeta=7.81\times10^{-6}$}}}%
    \put(0,0){\includegraphics[width=\unitlength,page=8]{ASym_bif.pdf}}%
    \put(0.48981593,0.0022788){\color[rgb]{0,0,0}\makebox(0,0)[lb]{\smash{$\frac{F-F_{\mbox{min}}}{F_{\mbox{min}}}$}}}%
    \put(0.020205,0.36827042){\color[rgb]{0,0,0}\makebox(0,0)[lb]{\smash{$\theta$}}}%
  \end{picture}%
\endgroup%
}
\caption{The perfect rigid link lattice acts as a limit for the continuum lattice with decreasing imperfections. Above shows the postbuckling behaviour of the continuum lattice found through FEM simulations for decreasing initial imperfection size in red for a mode with translational symmetry (left, $a=0.945, b=0.92$) and mirror symmetry (right, $a=0.96, b=0.92$), while the black curves show the postbuckling behaviour for a lattice of the same geometry predicted through Eq.~(\ref{bif}) with appropriate parameters of $\tau$ and $\kappa_n$. It is noted that the value of $F_c$ used in Eqs.~(\ref{bif}) is taken from FEM simulations to correct for the error shown in figure \ref{linear}. Other parameters used are given in caption of figure \ref{mode_diagrams}. Imperfections of varying magnitude were added to the structure in the from on the first eigenmode as predicted through the linear studies. In the case of the translationally symmetric mode (left) $\zeta$ represents the initial value of $\theta_1$, in the mirror symmetric case (right), $\zeta$ gives the initial value of $\theta$.\label{bifurcations}}
\end{center}
\end{figure}

The postbuckling of the rigid lattice can also be described analytically. Beyond the buckling threshold, due to the incompressibility of the links, the system remains in a configuration closely approximated by the modes predicted by the linear analysis. The magnitude of these modes can be predicted through energy considerations. Substituting Eq.~(\ref{mode_sym}) or Eqs.~(\ref{mode_asym_a} \& \ref{mode_asym_b}) into Eq.~(\ref{energy}), for the translationally symmetric or mirror symmetric mode respectively, we find that in both cases the energy of the system can be expressed as, 
\begin{equation}
U \approx (\alpha_i - F\beta_i) B^2 + \zeta_i F B^4,
\end{equation}
where $B$ characterises the magnitude deformation present within the system (in the case of a translationally symmetric mode, $B = A$ from Eq.~(\ref{mode_sym}), while for the mode with mirror symmetry, $B = \theta$ from Eq.~(\ref{mode_asym_a})), $\alpha_i, \beta_i$ and $\zeta_i$ are constants that take values dependent on the mode being investigated. 
It can then be shown that the minimum energy configuration is realised as 
\begin{align}
B = \begin{cases} 0 \quad &\mbox{for} \quad F < F_{\tiny \mbox{min}}\\ \sqrt{\frac{\alpha_i-F\beta_i}{2\zeta_iF}} \quad &\mbox{for} \quad F > F_{\mbox{\tiny min}} \end{cases}\label{bif}
\end{align}
Thus, considering Eqs.~(\ref{kappa} and \ref{tau}), we make predictions about the nature of the postbuckling behaviour of the continuum lattice presented in figure \ref{Systems}. 
These predictions, alongside the postbuckling behaviour found through finite element simulations are shown in figure \ref{bifurcations}, where it is noted that the rigid link analysis appears as a limit to which the continuum lattice converges in the limit of decreasing magnitudes of imperfections in the system. 
In figure \ref{Systems}, the value of $F_{\mbox{\tiny min}}$ used in Eq.~(\ref{bif}) has been taken from the finite element simulations to decrease the error (these errors can be seen in figure \ref{linear}). Nonetheless, the functional form of the graph shows good agreement between the two methods.

\subsection{Auxetic metamaterials}
\begin{figure}
\begin{center}
\resizebox{6cm}{!}{\begingroup%
  \makeatletter%
  \providecommand\color[2][]{%
    \errmessage{(Inkscape) Color is used for the text in Inkscape, but the package 'color.sty' is not loaded}%
    \renewcommand\color[2][]{}%
  }%
  \providecommand\transparent[1]{%
    \errmessage{(Inkscape) Transparency is used (non-zero) for the text in Inkscape, but the package 'transparent.sty' is not loaded}%
    \renewcommand\transparent[1]{}%
  }%
  \providecommand\rotatebox[2]{#2}%
  \ifx\svgwidth\undefined%
    \setlength{\unitlength}{256.00000096bp}%
    \ifx\svgscale\undefined%
      \relax%
    \else%
      \setlength{\unitlength}{\unitlength * \real{\svgscale}}%
    \fi%
  \else%
    \setlength{\unitlength}{\svgwidth}%
  \fi%
  \global\let\svgwidth\undefined%
  \global\let\svgscale\undefined%
  \makeatother%
  \begin{picture}(1,0.67842675)%
    \put(0,0){\includegraphics[width=\unitlength,page=1]{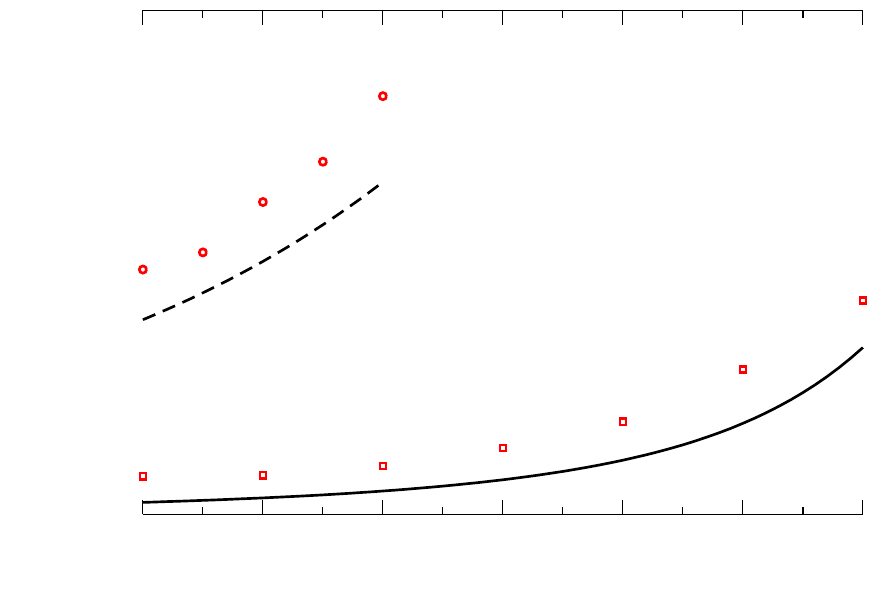}}%
    \put(0.14030313,0.05224513){\color[rgb]{0,0,0}\makebox(0,0)[lb]{\smash{0.9}}}%
    \put(0.26787905,0.05224513){\color[rgb]{0,0,0}\makebox(0,0)[lb]{\smash{0.91}}}%
    \put(0.401449,0.05224513){\color[rgb]{0,0,0}\makebox(0,0)[lb]{\smash{0.92}}}%
    \put(0.53720597,0.05224513){\color[rgb]{0,0,0}\makebox(0,0)[lb]{\smash{0.93}}}%
    \put(0.67142393,0.05224513){\color[rgb]{0,0,0}\makebox(0,0)[lb]{\smash{0.94}}}%
    \put(0.80701889,0.05192112){\color[rgb]{0,0,0}\makebox(0,0)[lb]{\smash{0.95}}}%
    \put(0.94156086,0.05224513){\color[rgb]{0,0,0}\makebox(0,0)[lb]{\smash{0.96}}}%
    \put(0.53688197,0.00350778){\color[rgb]{0,0,0}\makebox(0,0)[lb]{\smash{$b$}}}%
    \put(0,0){\includegraphics[width=\unitlength,page=2]{Poisson.pdf}}%
    \put(0.09566584,0.08805036){\color[rgb]{0,0,0}\makebox(0,0)[lb]{\smash{-1}}}%
    \put(0.06804465,0.20177519){\color[rgb]{0,0,0}\makebox(0,0)[lb]{\smash{-0.9}}}%
    \put(0.06836865,0.315014){\color[rgb]{0,0,0}\makebox(0,0)[lb]{\smash{-0.8}}}%
    \put(0.06836865,0.42841482){\color[rgb]{0,0,0}\makebox(0,0)[lb]{\smash{-0.7}}}%
    \put(0.06772065,0.54181564){\color[rgb]{0,0,0}\makebox(0,0)[lb]{\smash{-0.6}}}%
    \put(0.06869266,0.65505446){\color[rgb]{0,0,0}\makebox(0,0)[lb]{\smash{-0.5}}}%
    \put(-0.00174421,0.37138027){\color[rgb]{0,0,0}\makebox(0,0)[lb]{\smash{$\nu$}}}%
    \put(0,0){\includegraphics[width=\unitlength,page=3]{Poisson.pdf}}%
    \put(0.63480911,0.54777274){\color[rgb]{0,0,0}\makebox(0,0)[lb]{\smash{$a = 0.96$}}}%
    \put(0,0){\includegraphics[width=\unitlength,page=4]{Poisson.pdf}}%
    \put(0.63480911,0.50516346){\color[rgb]{0,0,0}\makebox(0,0)[lb]{\smash{$a = 0.985$}}}%
    \put(0,0){\includegraphics[width=\unitlength,page=5]{Poisson.pdf}}%
    \put(0.63480911,0.46247317){\color[rgb]{0,0,0}\makebox(0,0)[lb]{\smash{FEM, $a=0.96$}}}%
    \put(0,0){\includegraphics[width=\unitlength,page=6]{Poisson.pdf}}%
    \put(0.63480911,0.41856787){\color[rgb]{0,0,0}\makebox(0,0)[lb]{\smash{FEM, $a = 0.985$}}}%
    \put(0,0){\includegraphics[width=\unitlength,page=7]{Poisson.pdf}}%
  \end{picture}%
\endgroup%
}
\caption{Points show the minimum Poisson's ratio exhibited by the continuum lattice for a lattice with $N_x = 10$, $N_y = 9$, $L = 20$mm and $l_c = 3.75$mm (results obtained by FEM). The black curves show the Poisson's ratio of the analagous rigid link lattice. Increasingly good agreement is found for increasing $a$. \label{Poisson_ratio}}
\end{center}
\end{figure}

The antisymmetric mode of this system is strongly associated with auxetic behaviour of the lattice~\cite{Buckling_aux1, Buckling_aux2, Buckling_aux3}, where the Poisson's ratio of the structure is a function of the loading parameter \cite{Buckling_aux1}. 
We stress that in the limit of large deformations, in many cases, this dependence has a well defined limit (as investigated in \cite{Buckling_aux1}). By utilising the rigid link system, we are able to obtain estimates for this limit. For the geometry investigated here, it is found that, for large strains, localisation of deformations occur close to the boundaries. In figure \ref{Poisson_ratio} we plot the minimum value of the Poisson's ratio observed in FEM simulations (observed immediately before the localisation of deformation) and the Poisson's ratio predicted the rigid link analysis for various values of $a$ and $b$ and for the parameters of the system are given in the caption of figure \ref{mode_diagrams}. 
\section{Discussion}
In this paper, we have elucidated the fundamental mechanisms behind the auxetic behaviour of a broad class lattice based metamaterials, and presented a new methodology in the analysis of such materials. We have applied symmetry arguments to simplify the problem before finding fully analytic solutions for the behaviour of the rigid link lattice. We have shown excellent agreement between the rigid link mechanism results and the mechanical response of buckling lattices made up of a soft isotropic material, encapsulating both the linear stability studies and post-buckling behaviour. While in this work we have focused on the auxetic properties of the lattices, the methodology is well suited to the analysis of materials exhibiting frustrated mechanics and hysteretic behaviour. We hypothesise that the methodology presented here can be a useful tool in the design of structures with the mechanical response programmed into the geometry of the material. 

\section{Appendix: FEM methods}
Finite element studies were undertaken using COMSOL Multiphysics 5.2 \cite{COMSOL}. Both the linear and stationary studies were preformed on the same mesh. The mesh density varied a depending on the aspect ratio of the members considered, however approximately 200 - 1500 mesh elements were used per unit cell. Mesh refinement studies were undertaken to check for convergence of results. In all simulations, a Young's Modulus of 170MPa and Poisson's ratio of 0.25 has been used.  

We used both linear buckling analyses for figure \ref{linear} and stationary studies for figure \ref{bifurcations}. In both cases, the boundary conditions considered here were such that the upper and lower boundaries of the lattice were permitted to move freely in the $x$ direction, while the left and right boundaries were premitted to translate freely in any direction. The upper, lower, left and right boundaries were not permitted to rotate. 
In the linear buckling studies, an applied load was applied to the upper and lower boundaries and the linear buckling load was found. In the post buckling studies, the upper boundaries had a displacement in the $y$ direction imposed while the lower boundaries were fixed in the $y$ direction. The reaction forces on the upper and lower boundaries were then measured. For the stationary studies, small imperfections in the shape of the first eigenmode (found through the linear buckling studies) were then added to the mesh through the use of MeshPerturb 1.0 \cite{MeshPerturb}.

For the measurements of the Poisson's ratio, the displacement of the central points of the nodes within the structure were recorded. These displacements were then analysed to find the Poisson's ratio, the displacements of a single unit cell towards the centre of the lattice were taken to be representative of the structure.

\section{Funding}
DRK acknowledges funding support from Academy of Finland and Aalto Science Institute. CZ acknowledges funding from Aalto Science Institute. LT acknowledges funding from Aalto Science Institute thematic program ``Challenges in large geometric structures and big data''.


\end{document}